# Isotropic MIMO Interference Channels without CSIT: The Loss of Degrees of Freedom

Yan Zhu and Dongning Guo

Dept. of EECS Northwestern University
Evanston, IL 60208 United States

***Abstract*— This paper studies two-user MIMO interference channel with isotropic fading. We assume that users are equipped with arbitrary number of antennas and the channel state information (CSI) is available at receivers only. An outer bound is obtained for the degree of freedom region, which suggests the loss of degrees of freedom due to the lack of CSI at transmitters under many circumstances.**

## I. Introduction

An important advance toward the understanding of interference channels is the characterization of the capacity of the two-user Gaussian interference channel within one bit [1]. Since then new results have been obtained for $K$-user interference channels [2]–[5] and MIMO interference channels [6]–[8].

Several schemes have been developed to show the achievability in those works. For example, in one-antenna two-user scenario, results in [1] imply that the simplified Han-Kobayashi (HK) scheme [9] by setting the reception power of private message at non-intended user right below noise level is nearly optimal. However, HK scheme with Gaussian random coding is not enough for more than two users [5]. Instead, interference alignment at either signal level [2] or at code level [3]–[5], is shown to achieve the maximum degrees of freedom (DoF). For the system with multi-antenna, interference alignment also plays an important role [10].

It is important to note that all the above coding schemes rely on perfect knowledge of channel state information (CSI) at the transmitters. In fact, the capacity region of the system can be highly sensitive to the channel coefficients [3]. Furthermore, assuming full CSI is known at both transmitters and receivers may not be practical. There have been a few limited studies on the scenario with lack of CSI at transmitters. For example, compound interference channel is studied in [11], diversity-multiplexity tradeoff (DMT) of 2-user one-antenna interference channel under slow fading is studied in [12] [13], the DoF region for two special cases under Rayleigh fast fading is established in more recent work [14].

In this paper, we consider the fast fading channel without CSI at transmitters and study the degrees of freedom in ergodic sense. An outer bound of DoFs with isotropic fading for systems equipped with arbitrary numbers of antennas is established, which can be consider as a generalization of corresponding results in [14] and implies a substantial DoF loss due to the lack of CSI at transmitters under many circumstances.

This work has been supported by NSF under grant CCF-0644344.

## II. System Model and Problem Formulation

Consider a two-user interference channel, where each user has dedicated information for its receiver. Suppose that transmitter $i$ $(= 1, 2)$ is equipped with $M_i$ transmit antennas and receiver $i$ is equipped with $N_i$ receive antennas. Formally, we have

$$\boldsymbol{y}_1[m] = \mathbf{H}_{11}[m]\boldsymbol{x}_1[m] + \mathbf{H}_{21}[m]\boldsymbol{x}_2[m] + \boldsymbol{z}_1[m] \quad (1a)$$

$$\boldsymbol{y}_2[m] = \mathbf{H}_{12}[m]\boldsymbol{x}_1[m] + \mathbf{H}_{22}[m]\boldsymbol{x}_2[m] + \boldsymbol{z}_2[m] \quad (1b)$$

where $m$ is the time index, $\boldsymbol{x}_i(M_i \times 1)$ denotes the transmit signal of user $i$, $\mathbf{H}_{ij}(N_j \times M_i)$ denotes the channel matrix between transmitter $i$ and receiver $j$, and $\boldsymbol{z}_i(N_i \times 1) \sim \mathcal{CN}(0, \sigma^2 \mathrm{I})$ denotes the thermal noise at receiver $i$, which consists of independent identically distributed (i.i.d.) circularly symmetric complex Gaussian (CSCG) entries. The fading processes $\{\mathbf{H}_{ij}[m]\}$ and $\{\boldsymbol{z}_i[m]\}$ $(i, j = 1, 2)$ are all i.i.d. over time and mutually independent. For each user, total transmit power is no greater than $P$, *i.e.*,

$$\frac{1}{n}\sum_{m=1}^{n} \|\boldsymbol{x}_i[m]\|^2 \leq P\,. \quad (2)$$

Furthermore, we assume that each realization of $\mathbf{H}_{ij}(i = 1, 2)$ is available at receiver $j$ only, while transmitters have no knowledge about channel state except for their statistics.

*Definition 1:* Complex random matrix $\mathbf{R}$ is isotropic if and only if for every unitary matrix $\mathrm{Q}$, the random matrices $\mathbf{R}$ and $\mathbf{R}\mathrm{Q}$ are identically distributed, which we denote by $\mathbf{R} \sim \mathbf{R}\mathrm{Q}$.

In this paper, we consider the situation where each $\mathbf{H}_{ij}$ is isotropic, almost surely full rank, and of finite average power, *i.e.*, $\mathbb{E}\|\mathbf{H}_{ij}\|^2 < \infty$. Note that many important fading models fall into this category, including, in particular, Raleigh fading, where the elements of the channel matrices are i.i.d. CSCG random variables.

Following the information theoretic convention, we say the rate pair $(R_1, R_2)$ is achievable if for each user $i$, there exists an $n$-length codebook with size $\lceil 2^{nR_i} \rceil$ such that the average decoding error at both receivers vanishes. We define the degrees of freedom region as

$$\mathcal{D} = \Big\{ (d_1, d_2) \Big| \exists \text{ achievable pair } (R_1(P), R_2(P)) \text{ s.t.}$$

$$(d_1, d_2) = \lim_{P\to\infty} \frac{1}{\log(1 + P/\sigma^2)} (R_1(P), R_2(P))\Big\}.$$

Note that $\log(1 + P/\sigma^2)$ is the capacity of point-to-point channel with single antenna. Therefore, $d_i$ is the asymptotic number of point-to-point link capacity user $i$ can achieve. Although the DoF is still away from capacity region characterization, it is an important metric for communication systems with high signal-to-noise ratio (SNR) [15] [10] [16].

## III. Main Results

### A. The Outer Bound of DoF Region

*Theorem 1:* The non-negative DoF pair $(d_1, d_2)$ which is achievable for channel (1) with full rank isotropic fading, must satisfy the following constraints

$$d_i \le \min(M_i, N_i), \quad i = 1, 2 \quad (3a)$$

$$d_1 + \frac{\min(N_1, N_2, M_2)}{\min(N_2, M_2)} d_2 \le \min(M_1 + M_2, N_1) \quad (3b)$$

$$\frac{\min(N_1, N_2, M_1)}{\min(N_1, M_1)} d_1 + d_2 \le \min(M_1 + M_2, N_2) \quad (3c)$$

The proof is relegated to Section IV.

Note that in point-to-point MIMO systems, DoF is given by $\min(M, N)$ [15], where $M$ and $N$ are the numbers of transmit antennas and receive antennas, respectively. Therefore, (3a) simply says that DoFs of user $i$ can not exceed the maximal value when the other user is absent, which is trivial. To investigate (3b) and (3c), we assume that user 2 has no less receive antennas than user 1 ($N_2 \ge N_1$). Therefore, (3b) and (3c) can be simplified as

$$d_1 + \frac{\min(N_1, M_2)}{\min(N_2, M_2)} d_2 \le \min(M_1 + M_2, N_1) \quad (4)$$

$$d_1 + d_2 \le \min(M_1 + M_2, N_2). \quad (5)$$

It is shown in Appendix I that (5) is redundant with respect to (4). And (4) says that if the DoFs of user 2 is weighed by $\min(N_1, M_2)/\min(N_2, M_2)$, the weighed sum DoFs is limited by the total DoFs of two-user multiple access channel (MAC) with $M_i$ transmit antennas for user $i$ $(= 1, 2)$ and $N_1$ receive antennas. Due to the symmetry between (3b) and (3c), for case where $N_1 \ge N_2$, similar discussion can be conducted by reversing the roles of user 1 and user 2 in (4) and (5). Assuming $N_2 \ge N_1$, Fig. 1 illustrates the outer bound of the DoF regions for different cases.

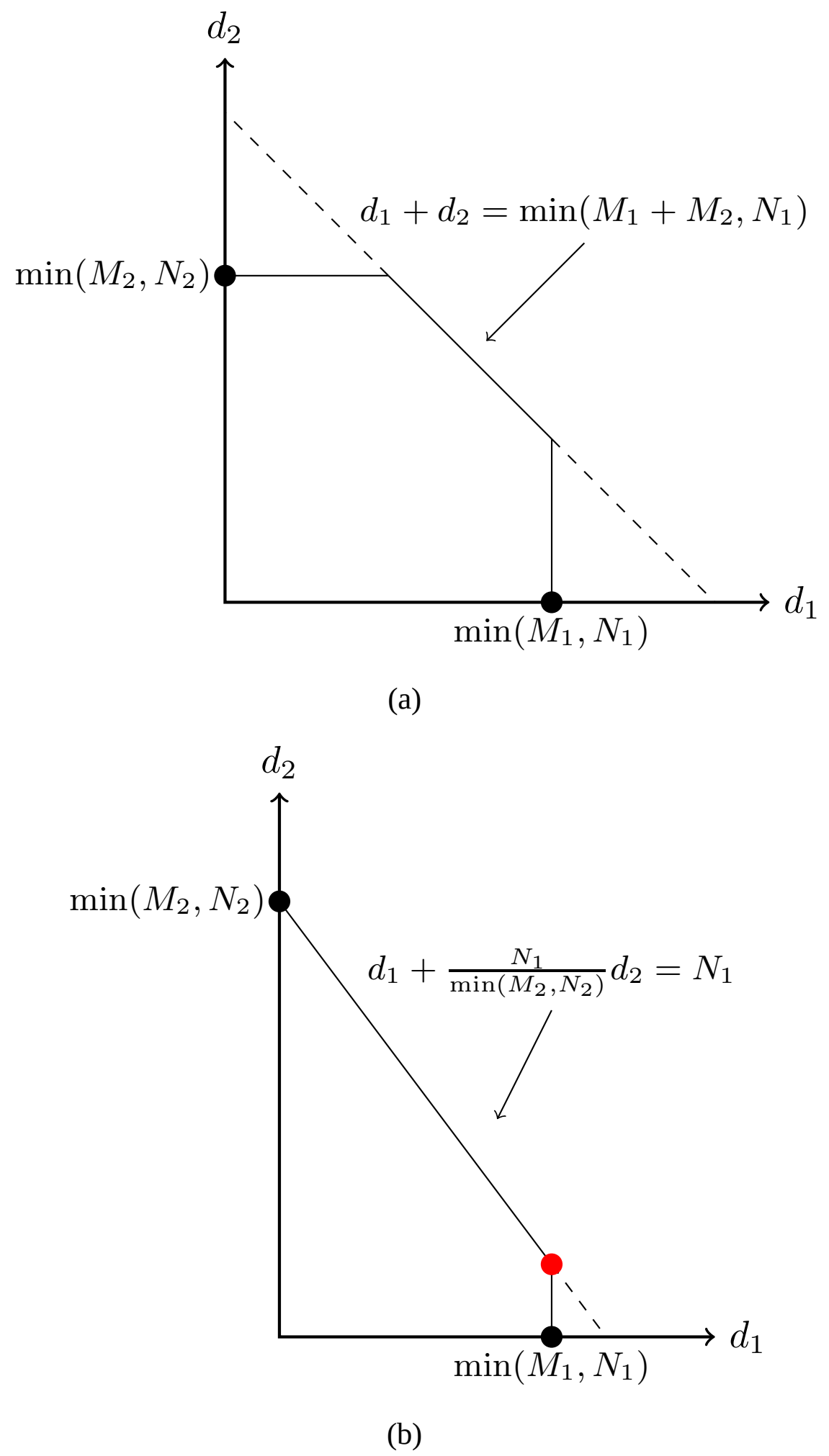

Fig. 1. Outer bound of DoF regions for different cases. (a) $N_2 \ge N_1 \ge M_2$ (b) $N_2 \ge N_1$ and $M_2 > N_1$

The general achievable schemes of the outer bound given by Theorem 1 are yet to be found. However, the optimality can be shown for some special cases.

It is shown in [14, Theorem 2] that the DoF region for Rayleigh fading with $N_1 \ge M_2$ and $N_2 \ge M_1$ is

$$\left\{ (d_1, d_2) \in \mathbb{R}_+^2 \;\middle|\; \begin{array}{c} d_i \le \min(M_i, N_i) \quad i = 1, 2 \\ d_1 + d_2 \le \min(N_1, N_2) \end{array} \right\}, \quad (6)$$

which can be obtained from Theorem 1 for general isotropic fading. By assuming further $N_2 \ge N_1$, in addition to the single user bounds (3a), we have $d_1 + d_2 \le \min(M_1 + M_2, N_1)$; by assuming $N_1 \ge N_2$, we have $d_1 + d_2 \le \min(M_1 + M_2, N_2)$. And in the both sub-cases, allowing both receivers being able to decode both users' information, we can achieve the outer bound. Therefore, the DoF region is

$$\left\{ (d_1, d_2) \in \mathbb{R}_+^2 \;\middle|\; \begin{array}{c} d_i \le \min(M_i, N_i) \quad i = 1, 2 \\ d_1 + d_2 \le \min(M_1 + M_2, N_1, N_2) \end{array} \right\}. \quad (7)$$

By noting the fact that the sum constraint in (7) is redundant when $M_1 + M_2 \le N_1, N_2$, we see that (6) and (7) are equivalent.

Another special case is that $M_1 \ge N_1$ and $M_2 \ge N_2$. Applying Theorem 1 with further assumption of $N_2 \ge N_1$, (3a) and (3c) are redundant, and (3b) can be simplified to $d_1/N_1 + d_2/N_2 \le 1$; applying Theorem 1 with further of assumption $N_2 \le N_1$, (3a) and (3b) are redundant, and (3c) can be simplified to $d_1/N_1 + d_2/N_2 \le 1$. And for both sub-cases, the outer bound can be achieved by time division multiple access (TDMA). Therefore, the DoF region for $M_1 \ge N_1$ and $M_2 \ge N_2$ is $\{(d_1, d_2) \in \mathbb{R}_+^2 | d_1/N_1 + d_2/N_2 \le 1\}$, which generalize the known results in [14, Theorem 2] to the isotropic fading MIMO interference channel.

## B. System with CSI

The DoFs for MIMO interference channels with full CSI at all transmitters and receivers is studied in [10], where the total/sum DoF is shown to be $\min(M_1 + M_2, N_1 + N_2, \max(M_1, N_2), \max(M_2, N_1))$. Technique used in [10] is based on so called interference alignment. Roughly speaking, transmitters design their transmit signal based on the CSI such that they can minimize the interference at both receivers. Therefore, to exploit interference alignment, CSI plays a central role. However, when the CSI is not known to the transmitters, the two users can not align the interference.

Indeed, without transmitter-side CSI, there can be substantial loss of DoF under many circumstances. The absolute capacity loss can be arbitrarily large as SNR increases. For example, consider a symmetric system, where $M_i = M$ and $N_i = N$. With full CSI, the total DoF is $\min(2M, 2N, \max(M, N))$; while with receiver CSI only, the total DoF is $\min(2M, N)$. Thus, the DoF loss is $\min(N, (M-N)^+)$, where $(x)^+ := \max(0, x)$. Therefore, when $N \geq M$, there is no DoF loss, which means that abundance of receive antennas enables each receiver to separate interference and desired signal without exploiting interference alignment at transmit sides; when $N < M$, the DoF loss is strictly positive, which means that the lack of receiver antennas makes the interference alignment at transmitters indispensable. An interesting implication for system design is: While the DoF is limited by the party with fewer antennas, the achievability depends on the party with more antennas.

# IV. Proof

In this section, we prove Theorem 1. The assumption of isotropic fading is crucial to the development. We first summarize the properties of isotropic random matrix in the following lemma.

*Lemma 1:* Let $\mathbf{R}(N \times M)$ be an isotropic complex random matrix and define $K = \min(M, N)$. Then there exists a decomposition $\mathbf{R} \sim \mathbf{W}\mathbf{\Lambda}\mathbf{V}^\dagger$ such that $\mathbf{W}$, $\mathbf{\Lambda}$ and $\mathbf{V}$ have following properties:

1) $\mathbf{W}_{N\times K}$ and $\mathbf{V}_{M\times K}$ are random matrices satisfying $\mathbf{V}^\dagger\mathbf{V} = \mathbf{W}^\dagger\mathbf{W} = \mathrm{I}$, and $\mathbf{\Lambda}_{K\times K}$ is a diagonal random matrix with non-negative diagonal elements.
2) $\mathbf{V}$ is independent of $(\mathbf{W}, \mathbf{\Lambda})$ and is uniformly distributed on $\mathcal{V}$, where $\mathcal{V} = \{\mathrm{V}_{M\times K} : \mathrm{V}^\dagger\mathrm{V} = \mathrm{I}\}$.

*Proof:* For any realization of $\mathbf{R}$, we have singular value decomposition (SVD). Therefore, $\mathbf{R} = \mathbf{W}\mathbf{\Lambda}\mathbf{V}_1^\dagger$. By the definition of SVD, property 1 holds. The remaining is to replace $\mathbf{V}_1$ with $\mathbf{V}$, which fulfills the property 2.

Let $\mathcal{Q} = \{\mathrm{Q}_{M\times M} : \mathrm{Q}^\dagger\mathrm{Q} = \mathrm{I}\}$. Choose a random matrix $\mathbf{Q}$ such that it is independent of $\mathbf{R}$ and has a uniform distribution on $\mathcal{Q}$. Set $\mathbf{V} = \mathbf{Q}^\dagger\mathbf{V}_1$. Given $\mathbf{Q} = \mathrm{Q}$, $\mathbf{R} \sim \mathbf{R}\mathrm{Q}$. Since $\mathbf{Q}$ is uniform, we have $\mathbf{R} \sim \mathbf{R}\mathbf{Q}$, thus, $\mathbf{R} \sim \mathbf{W}\mathbf{\Lambda}\mathbf{V}^\dagger$. Furthermore, given $\mathbf{V}_1 = \mathrm{V}_1$, $\mathbf{V} = \mathbf{Q}^\dagger\mathrm{V}_1$ can traverse all the points in $\mathcal{V}$. Again, because $\mathbf{Q}$ has a uniform distribution, the density function $f_{\mathbf{V}|\mathbf{V}_1}(\mathrm{V}; \mathrm{V}_1)$ is a constant on $\mathcal{V}$. Therefore, $\mathbf{V}$ is uniform and independent of $\mathbf{V}_1$. Note that given $\mathbf{V}_1$, $\mathbf{V}$ is independent of $(\mathbf{W}, \mathbf{\Lambda})$, which completes the proof. ■

Intuitively, Lemma 1 suggests that we can decompose the isotropic matrix into "amplitude" ($\mathbf{\Lambda}$) and "phase" ($\mathbf{V}$), which are mutually independent.

Because (3a) is trivial and (3b) and (3c) are symmetry, it is sufficient to show (4) with assumption that $N_2 \geq N_1$. Our proof is inspired by "Marton-like" expansion used in [17] and the generalized supper-additivity of differential entropy [18]. In this work, we use the mutual information in lieu of differential entropy for better insights, since the goal is basically to upper bound the total mutual information.

We use following conventions: $\boldsymbol{u}_{k,i}^l$ denotes the sequence of random variables, vectors, or matrices $\{\boldsymbol{u}_i[k], \boldsymbol{u}_i[k+1], \ldots \boldsymbol{u}_i[l]\}$. For simplicity, we use $\mathbf{H}^n$ to denote all the channel state information from instant 1 to $n$, *i.e.*, $\mathbf{H}^n = \{\mathbf{H}_{1,ij}^n\}_{i,j=1,2}$.

Now, consider the following bounds, by revealing $\boldsymbol{x}_1$ to user 2,

$$R_1 \leq \frac{1}{n} I(\boldsymbol{y}_{1,1}^n; \boldsymbol{x}_{1,1}^n | \mathbf{H}^n) + \epsilon(n) \tag{8}$$

$$R_2 \leq \frac{1}{n} I(\boldsymbol{y}_{1,2}^n; \boldsymbol{x}_{1,2}^n | \boldsymbol{x}_{1,1}^n, \mathbf{H}^n) + \epsilon(n)\,. \tag{9}$$

Note that (8) and (9) are due to the Fano's inequality, where $\epsilon(n)$ vanishes if the error probability is required to vanish. Recall that the right hand of (4) is total DoF of 2-user MAC channel, which suggests that we view (8) as

$$R_1 \leq \frac{1}{n} I(\boldsymbol{y}_{1,1}^n; \boldsymbol{x}_{1,1}^n, \boldsymbol{x}_{1,2}^n | \mathbf{H}^n) - \frac{1}{n} I(\boldsymbol{y}_{1,1}^n; \boldsymbol{x}_{1,2}^n | \boldsymbol{x}_{1,1}^n, \mathbf{H}^n) + \epsilon(n)$$

Let $\boldsymbol{s}_1 = \mathbf{H}_{21}\boldsymbol{x}_2 + \boldsymbol{z}_1$ and $\boldsymbol{s}_2 = \mathbf{H}_{22}\boldsymbol{x}_2 + \boldsymbol{z}_2$. Then (8) and (9) can be rewritten as

$$R_1 \leq \frac{1}{n} I(\boldsymbol{y}_{1,1}^n; \boldsymbol{x}_{1,1}^n, \boldsymbol{x}_{1,2}^n | \mathbf{H}^n) - \frac{1}{n} I(\boldsymbol{s}_{1,1}^n; \boldsymbol{x}_{1,2}^n | \mathbf{H}^n) + \epsilon(n) \tag{10}$$

$$R_2 \leq \frac{1}{n} I(\boldsymbol{s}_{1,2}^n; \boldsymbol{x}_{1,2}^n | \mathbf{H}^n) + \epsilon(n)\,. \tag{11}$$

Note that signals $\boldsymbol{s}_1$ and $\boldsymbol{s}_2$ both come from the second user. To obtain the upper bound, one needs to bound difference between these two mutual informations with proper weights. To obtain that, we need some lemmas.

As we mentioned before, we can separate the isotropic fading effect into "amplitude" part and "phase" part. Following lemma says that if we change the statistics of "amplitude", the information loss can be bounded below.

*Lemma 2:* Let $\mathbf{\Lambda}_1$ and $\mathbf{\Lambda}_2$ be two $N$-by-$N$ diagonal random matrices with positive diagonal elements almost surely, which both are independent of a random vector $\boldsymbol{x}$ and a CSCG random vector $\boldsymbol{z}$. Define random matrix $\mathbf{\Lambda}_{\min} = \min(\mathbf{\Lambda}_1, \mathbf{\Lambda}_2)$, where $\min$ takes element-wisely. Then we have

$$I(\mathbf{\Lambda}_1\boldsymbol{x} + \boldsymbol{z}; \boldsymbol{x}|\mathbf{\Lambda}_1) \geq I(\mathbf{\Lambda}_2\boldsymbol{x} + \boldsymbol{z}; \boldsymbol{x}|\mathbf{\Lambda}_2) - \mathbb{E}\log^+\det\mathbf{\Lambda}_2 - \mathbb{E}\left[\log^+\frac{1}{\det\mathbf{\Lambda}_{\min}}\right], \tag{12}$$

where $\log^+(x) := \max(0, \log x)$.

*Proof:* Define $\Lambda_{\min} = \min(\Lambda_1, \Lambda_2)$, where $\Lambda_1$ and $\Lambda_2$ are realizations of $\mathbf{\Lambda}_1$ and $\mathbf{\Lambda}_2$, respectively. We know that $\Lambda_{\min} \preceq \Lambda_1, \Lambda_2$. It suffices to show that for each realization of $\mathbf{\Lambda}_1$ and $\mathbf{\Lambda}_2$, we have

$$I(\Lambda_1 \boldsymbol{x} + \boldsymbol{z}; \boldsymbol{x}) - I(\Lambda_2 \boldsymbol{x} + \boldsymbol{z}; \boldsymbol{x}) \geq -\log^+(\det \Lambda_2) - \log^+\left(\frac{1}{\det \Lambda_{\min}}\right). \quad (13)$$

Let $\boldsymbol{z}'$ be a random vector independent of $\boldsymbol{z}$ but with same distribution. By data process inequality [19],

$$\begin{aligned}
&I(\Lambda_1 \boldsymbol{x} + \boldsymbol{z}; \boldsymbol{x}) - I(\Lambda_2 \boldsymbol{x} + \boldsymbol{z}; \boldsymbol{x}) \\
&\quad \geq I(\Lambda_{\min} \boldsymbol{x} + \boldsymbol{z}; \boldsymbol{x}) - I(\Lambda_2 \boldsymbol{x} + \boldsymbol{z}; \boldsymbol{x}) \\
&\quad = I(\Lambda_2 \Lambda_{\min}^{-1}(\Lambda_{\min} \boldsymbol{x} + \boldsymbol{z}); \boldsymbol{x}) - I(\Lambda_2 \boldsymbol{x} + \boldsymbol{z}; \boldsymbol{x}) \\
&\quad = I(\Lambda_2 \boldsymbol{x} + \Lambda_2 \Lambda_{\min}^{-1} \boldsymbol{z}; \boldsymbol{x}) - I(\Lambda_2 \boldsymbol{x} + \boldsymbol{z}; \boldsymbol{x}) \\
&\quad = I(\Lambda_2 \boldsymbol{x} + \boldsymbol{z} + (\Lambda_2 \Lambda_{\min}^{-1} - \mathrm{I})\boldsymbol{z}'; \boldsymbol{x}) - I(\Lambda_2 \boldsymbol{x} + \boldsymbol{z}; \boldsymbol{x}) \\
&\quad = -I(\Lambda_2 \boldsymbol{x} + \boldsymbol{z}; \boldsymbol{x} | \Lambda_2 \boldsymbol{x} + \boldsymbol{z} + (\Lambda_2 \Lambda_{\min}^{-1} - \mathrm{I})\boldsymbol{z}'). \quad (14)
\end{aligned}$$

Therefore, it boils down to upper bound the mutual information in (14), which can be rewritten as

$$\begin{aligned}
&I(\Lambda_2 \boldsymbol{x} + \boldsymbol{z}; \boldsymbol{z} + (\Lambda_2 \Lambda_{\min}^{-1} - \mathrm{I})\boldsymbol{z}' | \Lambda_2 \boldsymbol{x} + \boldsymbol{z} + (\Lambda_2 \Lambda_{\min}^{-1} - \mathrm{I})\boldsymbol{z}') \\
&\leq I(\boldsymbol{z}'; \boldsymbol{z} + (\Lambda_2 \Lambda_{\min}^{-1} - \mathrm{I})\boldsymbol{z}' | \Lambda_2 \boldsymbol{x} + \boldsymbol{z} + (\Lambda_2 \Lambda_{\min}^{-1} - \mathrm{I})\boldsymbol{z}'). \quad (15)
\end{aligned}$$

Since $\boldsymbol{z}' — \boldsymbol{z} + (\Lambda_2 \Lambda_{\min}^{-1} - \mathrm{I})\boldsymbol{z}' — \Lambda_2 \boldsymbol{x} + \boldsymbol{z} + (\Lambda_2 \Lambda_{\min}^{-1} - \mathrm{I})\boldsymbol{z}'$ form a Markov chain, (15) can be further upper bounded so that

$$\begin{aligned}
I(\Lambda_2 \boldsymbol{x} + \boldsymbol{z}; \boldsymbol{x} | \Lambda_2 \boldsymbol{x} + \boldsymbol{z} + (\Lambda_2 \Lambda_{\min}^{-1} - \mathrm{I})\boldsymbol{z}') &\leq I(\boldsymbol{z}'; \boldsymbol{z} + (\Lambda_2 \Lambda_{\min}^{-1} - \mathrm{I})\boldsymbol{z}') \\
&= \log\left(\frac{\det \Lambda_2}{\det \Lambda_{\min}}\right) \\
&\leq \log^+ \det \Lambda_2 + \log^+\left(\frac{1}{\det \Lambda_{\min}}\right).
\end{aligned}$$

We have thus establish (13). Lemma 2 follows by taking expectation on both sides. ■

To make use of the isotropic property of "phase," we introduce the following definition. In $\mathbb{C}^n$, we have $m$ positive numbers $p_1, \ldots, p_m$ and $m$ matrices $\mathrm{B}_1, \ldots, \mathrm{B}_m$, where $B_i$ is of $n_i \times n$ and $n_i \leq n$. We call the pair $(\{p_i\}_1^m, \{\mathrm{B}_i\}_1^m)$ *geometric* if $\mathrm{B}_i \mathrm{B}_i^\dagger = \mathrm{I}$, $i = 1, \ldots, m$ and $\sum_{i=1}^m p_i \mathrm{B}_i^\dagger \mathrm{B}_i = \mathrm{I}$. We have a generalized super-additive property of differential entropy [18], which is restated here in terms of mutual information.

*Theorem 2 (Carlen [18]):* Suppose $\boldsymbol{x}$ is a random vector in $\mathbb{C}^n$ and $\boldsymbol{z} \sim \mathcal{CN}(0, \sigma^2 \mathrm{I})$ is a CSCG vector of the same size as $\boldsymbol{x}$. If the pair $(\{p_i\}_1^m, \{\mathrm{B}_i\}_1^m)$ is geometric, then

$$\sum_{i=1}^m p_i I(\mathrm{B}_i(\boldsymbol{x} + \boldsymbol{z}); \boldsymbol{x}) \geq I(\boldsymbol{x} + \boldsymbol{z}; \boldsymbol{x}) \quad (16)$$

Theorem 2 has been shown in [18] using the Brascamp-Lieb inequalities. In Appendix II, we provide a simpler proof using the property of the minimum mean-square error (MMSE) and its relation with mutual information [20].

Now, we turn to the key lemma.

*Lemma 3:* Let $\mathbf{H}_1(N_1 \times M)$ and $\mathbf{H}_2(N_2 \times M)$ be two complex isotropic full rank random matrices where $N_1 \leq N_2$, $\boldsymbol{z}_1(N_1 \times 1)$ and $\boldsymbol{z}_2(N_2 \times 1) \sim \mathcal{CN}(0, \sigma^2 \mathrm{I})$ be two CSCG random vectors, and $\boldsymbol{x}$ be an $M \times 1$ random vector. All preceding random matrices and vectors are mutually independent. Then

$$\frac{\min(N_2, M)}{\min(N_1, M)} I(\mathbf{H}_1 \boldsymbol{x} + \boldsymbol{z}_1; \boldsymbol{x} | \mathbf{H}_1) \geq I(\mathbf{H}_2 \boldsymbol{x} + \boldsymbol{z}_2; \boldsymbol{x} | \mathbf{H}_2) + C \quad (17)$$

where $C$ is a constant, which does not depend on $\boldsymbol{x}$.

*Proof:* By Lemma 1, we have decompositions: $\mathbf{H}_1 \sim \mathbf{W}_1 \mathbf{\Lambda}_1 \mathbf{V}_1^\dagger$ and $\mathbf{H}_2 \sim \mathbf{W}_2 \mathbf{\Lambda}_2 \mathbf{V}_2^\dagger$, where $\mathbf{V}_i$ is independent of $(\mathbf{\Lambda}_i, \mathbf{W}_i)$. We first show that the value of mutual information $I(\mathbf{H}_i \boldsymbol{x} + \boldsymbol{z}_i; \boldsymbol{x} | \mathbf{H}_i)$ is independent of $\mathbf{W}_i$, and then bound the difference of the mutual informations in (17) using Lemma 2 and Theorem 2.

For $i = 1, 2$,

$$\begin{aligned}
I(\mathbf{H}_i \boldsymbol{x} + \boldsymbol{z}_i; \boldsymbol{x} | \mathbf{H}_i) &= I(\mathbf{W}_i \mathbf{\Lambda}_i \mathbf{V}_i^\dagger \boldsymbol{x} + \boldsymbol{z}_i; \boldsymbol{x} | \mathbf{W}_i, \mathbf{\Lambda}_i, \mathbf{V}_i) \\
&\stackrel{(a)}{=} I(\mathbf{\Lambda}_i \mathbf{V}_i^\dagger \boldsymbol{x} + \mathbf{W}_i^\dagger \boldsymbol{z}_i; \boldsymbol{x} | \mathbf{W}_i, \mathbf{\Lambda}_i, \mathbf{V}_i) \\
&\stackrel{(b)}{=} I(\mathbf{\Lambda}_i \mathbf{V}_i^\dagger \boldsymbol{x} + \boldsymbol{z}_i'; \boldsymbol{x} | \mathbf{\Lambda}_i, \mathbf{V}_i),
\end{aligned}$$

where $(a)$ follows from the fact that given $\mathbf{W}_i$, $\mathbf{\Lambda}_i \mathbf{V}_i^\dagger \boldsymbol{x} + \mathbf{W}_i^\dagger \boldsymbol{z}_i$ is a sufficient statistics of $\mathbf{W}_i \mathbf{\Lambda}_i \mathbf{V}_i^\dagger \boldsymbol{x} + \boldsymbol{z}_i$ for $\boldsymbol{x}$ [19], and in step $(b)$, we replace $\mathbf{W}_i^\dagger \boldsymbol{z}_i$ with $\boldsymbol{z}_i'(\min(M, N_i) \times 1)$, which is a CSCG random vector with covariance matrix $\mathbb{E}[\mathbf{W}_i^\dagger \boldsymbol{z}_i \boldsymbol{z}_i^\dagger \mathbf{W}_i | \mathbf{W}_i] = \sigma^2 \mathrm{I}$.

Define $\mathbf{\Lambda}_1' = \min(\mathbf{\Lambda}_1, \mathrm{I})$ and $\mathbf{\Lambda}_2' = \min(\mathbf{\Lambda}_2, \mathrm{I})$. Applying Lemma 2 with $\mathbf{\Lambda}_2$ set to I,

$$\begin{aligned}
&I(\mathbf{H}_1 \boldsymbol{x} + \boldsymbol{z}_1; \boldsymbol{x} | \mathbf{H}_1) \\
&\quad = I(\mathbf{\Lambda}_1 \mathbf{V}_1^\dagger \boldsymbol{x} + \boldsymbol{z}_1'; \boldsymbol{x} | \mathbf{\Lambda}_1, \mathbf{V}_1) \\
&\quad \geq I(\mathbf{V}_1^\dagger \boldsymbol{x} + \boldsymbol{z}_1'; \boldsymbol{x} | \mathbf{V}_1) - \mathbb{E}\left[\log^+ \frac{1}{\det(\mathbf{\Lambda}_1')}\right] \\
&\quad = \int I(\mathrm{V}_1^\dagger \boldsymbol{x} + \boldsymbol{z}_1'; \boldsymbol{x}) \mathsf{P}_{\mathbf{V}_1}(\mathsf{d}\mathrm{V}_1) - \mathbb{E}\left[\log^+ \frac{1}{\det(\mathbf{\Lambda}_1')}\right] \quad (18)
\end{aligned}$$

On the other hand, applying Lemma 2 to $I(\mathbf{H}_2 \boldsymbol{x} + \boldsymbol{z}_2; \boldsymbol{x} | \mathbf{H}_2)$ in the other direction with $\mathbf{\Lambda}_1$ set to I, we have

$$\begin{aligned}
&I(\mathbf{H}_2 \boldsymbol{x} + \boldsymbol{z}_2; \boldsymbol{x} | \mathbf{H}_2) \\
&\quad = I(\mathbf{\Lambda}_2 \mathbf{V}_2^\dagger \boldsymbol{x} + \boldsymbol{z}_2'; \boldsymbol{x} | \mathbf{\Lambda}_2, \mathbf{V}_2) \\
&\quad \leq I(\mathbf{V}_2^\dagger \boldsymbol{x} + \boldsymbol{z}_2'; \boldsymbol{x} | \mathbf{V}_2) + \mathbb{E} \log^+ \det \mathbf{\Lambda}_2 + \mathbb{E}\left[\log^+ \frac{1}{\det \mathbf{\Lambda}_2'}\right] \\
&\quad = \int I(\mathrm{V}_2^\dagger \boldsymbol{x} + \boldsymbol{z}_2'; \boldsymbol{x}) \mathsf{P}_{\mathbf{V}_2}(\mathsf{d}\mathrm{V}_2) + \mathbb{E} \log^+ \det \mathbf{\Lambda}_2 \\
&\qquad + \mathbb{E}\left[\log^+ \frac{1}{\det \mathbf{\Lambda}_2'}\right] \quad (19)
\end{aligned}$$

We need to compare the two integrals in (18) and (19) via Theorem 2. Let $m = \min(N_2, M)$ and $l = \min(N_1, M)$. Find an orthonormal basis in space $\mathbb{C}^m$, say, $\{u_i\}_1^m$; then construct $m$ subsets of $\{u_i\}_1^m$ such that each subset has $l$ elements and each $u_i$ is included in exact $l$ subsets; each subset corresponds to an $l \times m$ matrix, called $\mathrm{B}_1, \ldots, \mathrm{B}_m$. Apply Theorem 2 with $\mathrm{V}_2^\dagger \boldsymbol{x}$ as $\boldsymbol{x}$, and $p_1, \ldots, p_m = 1/l$. It is easy to check the pair $(\{p_i\}_1^m, \{\mathrm{B}_i\}_1^m)$ is geometric. Thus

$$I(\mathrm{V}_2^\dagger \boldsymbol{x} + \boldsymbol{z}_2'; \boldsymbol{x}) \le \sum_{i=1}^m \frac{1}{l} I(\mathrm{B}_i \mathrm{V}_2^\dagger \boldsymbol{x} + \boldsymbol{z}_1'; \boldsymbol{x}),$$

where we use the fact that $\mathrm{B}_i \boldsymbol{z}_2' \sim \boldsymbol{z}_1'$. Integrating with respect to $\mathsf{P}_{\mathbf{V}_2}(\mathsf{d}\mathrm{V}_2)$,

$$\begin{aligned}
&\int I(\mathrm{V}_2^\dagger \boldsymbol{x} + \boldsymbol{z}_2')\mathsf{P}_{\mathbf{V}_2}(\mathsf{d}\mathrm{V}_2) \\
&\le \sum_{i=1}^m \frac{1}{l} \int I(\mathrm{B}_i \mathrm{V}_2^\dagger \boldsymbol{x} + \boldsymbol{z}_1'; \boldsymbol{x})\mathsf{P}_{\mathbf{V}_2}(\mathsf{d}\mathrm{V}_2) \\
&= \sum_{i=1}^m \frac{1}{l} \int I(\mathrm{V}_1^\dagger \boldsymbol{x} + \boldsymbol{z}_1'; \boldsymbol{x})\mathsf{P}_{\mathbf{V}_1}(\mathsf{d}\mathrm{V}_1) \\
&= \frac{\min(N_2, M)}{\min(N_1, M)} \int I(\mathrm{V}_1^\dagger \boldsymbol{x} + \boldsymbol{z}_1'; \boldsymbol{x})\mathsf{P}_{\mathbf{V}_1}(\mathsf{d}\mathrm{V}_1) \qquad (20)
\end{aligned}$$

where we use the fact that $\mathrm{B}_i \mathbf{V}_2 \sim \mathbf{V}_1$.

Now substituting (20) into (19), we have

$$\begin{aligned}
&I(\mathbf{H}_2 \boldsymbol{x} + \boldsymbol{z}_2; \boldsymbol{x}|\mathbf{H}_2) \\
&\le \frac{\min(N_2, M)}{\min(N_1, M)} \int I(\mathrm{V}_1^\dagger \boldsymbol{x} + \boldsymbol{z}_1'; \boldsymbol{x})\mathsf{P}_{\mathbf{V}_1}(\mathsf{d}\mathrm{V}_1) \\
&\quad + \mathbb{E} \log^+ \det \mathbf{\Lambda}_2 + \mathbb{E}\left[\log^+ \frac{1}{\det \mathbf{\Lambda}_2'}\right]. \qquad (21)
\end{aligned}$$

Substituting (18) into (21), we obtain

$$\begin{aligned}
&I(\mathbf{H}_2 \boldsymbol{x} + \boldsymbol{z}_2; \boldsymbol{x}|\mathbf{H}_2) \\
&\le \frac{\min(N_2, M)}{\min(N_1, M)} I(\mathbf{H}_1 \boldsymbol{x} + \boldsymbol{z}_1; \boldsymbol{x}|\mathbf{H}_1) \\
&\quad + \frac{\min(N_2, M)}{\min(N_1, M)} \mathbb{E}\left[\log^+ \frac{1}{\det(\mathbf{\Lambda}_1')}\right] \\
&\quad + \mathbb{E} \log^+ \det \mathbf{\Lambda}_2 + \mathbb{E}\left[\log^+ \frac{1}{\det \mathbf{\Lambda}_2'}\right]. \qquad (22)
\end{aligned}$$

Note that the last three terms in (22) are constant independent of $\boldsymbol{x}$. Therefore, we complete the proof. ■

Go back to the proof of (4). We compare the two terms, $I(\boldsymbol{s}_{1,1}^n; \boldsymbol{x}_{1,2}^n|\mathbf{H}^n)$ and $I(\boldsymbol{s}_{1,2}^n; \boldsymbol{x}_{1,2}^n|\mathbf{H}^n)$ as follows

$$\begin{aligned}
&I(\boldsymbol{s}_{1,1}^n; \boldsymbol{x}_{1,2}^n|\mathbf{H}^n) - I(\boldsymbol{s}_{1,2}^n; \boldsymbol{x}_{1,2}^n|\mathbf{H}^n) \\
&\overset{(a)}{=} \sum_{i=1}^n \left\{ I(\boldsymbol{s}_{1,1}^i, \boldsymbol{s}_{i+1,2}^n; \boldsymbol{x}_{1,2}^n|\mathbf{H}^n) - I(\boldsymbol{s}_{1,1}^{i-1}, \boldsymbol{s}_{i,2}^n; \boldsymbol{x}_{1,2}^n|\mathbf{H}^n) \right\} \\
&\overset{(b)}{=} \sum_{i=1}^n \Big\{ I(\boldsymbol{s}_{1,1}^{i-1}, \boldsymbol{s}_{i+1,2}^n; \boldsymbol{x}_{1,2}^n|\mathbf{H}^n) \\
&\qquad + I(\boldsymbol{s}_{i,1}; \boldsymbol{x}_{1,2}^n|\boldsymbol{s}_{1,1}^{i-1}, \boldsymbol{s}_{i+1,2}^n, \mathbf{H}^n) \\
&\qquad - I(\boldsymbol{s}_{1,1}^{i-1}, \boldsymbol{s}_{i+1,2}^n; \boldsymbol{x}_{1,2}^n|\mathbf{H}^n) \\
&\qquad - I(\boldsymbol{s}_{i,2}; \boldsymbol{x}_{1,2}^n|\boldsymbol{s}_{1,1}^{i-1}, \boldsymbol{s}_{i+1,2}^n, \mathbf{H}^n) \Big\} \\
&= \sum_{i=1}^n \Big\{ I(\boldsymbol{s}_{i,1}; \boldsymbol{x}_{1,2}^n|\boldsymbol{s}_{1,1}^{i-1}, \boldsymbol{s}_{i+1,2}^n, \mathbf{H}^n) \\
&\qquad - I(\boldsymbol{s}_{i,2}; \boldsymbol{x}_{1,2}^n|\boldsymbol{s}_{1,1}^{i-1}, \boldsymbol{s}_{i+1,2}^n, \mathbf{H}^n) \Big\} \\
&\overset{(c)}{=} \sum_{i=1}^n \Big\{ I(\boldsymbol{s}_{i,1}; \boldsymbol{x}_{i,2}|\boldsymbol{s}_{1,1}^{i-1}, \boldsymbol{s}_{i+1,2}^n, \mathbf{H}^n) \\
&\qquad - I(\boldsymbol{s}_{i,2}; \boldsymbol{x}_{i,2}|\boldsymbol{s}_{1,1}^{i-1}, \boldsymbol{s}_{i+1,2}^n, \mathbf{H}^n) \Big\} \\
&\overset{(d)}{\ge} \sum_{i=1}^n \left( \frac{\min(N_1, M_2)}{\min(N_2, M_2)} - 1 \right) I(\boldsymbol{s}_{i,2}; \boldsymbol{x}_{i,2}|\boldsymbol{s}_{1,1}^{i-1}, \boldsymbol{s}_{i+1,2}^n, \mathbf{H}^n) \\
&\qquad + nC' \\
&\overset{(e)}{\ge} \sum_{i=1}^n \left( \frac{\min(N_1, M_2)}{\min(N_2, M_2)} - 1 \right) I(\boldsymbol{s}_{i,2}; \boldsymbol{x}_{i,2}|\boldsymbol{s}_{i+1,2}^n, \mathbf{H}^n) + nC' \\
&\overset{(f)}{=} \sum_{i=1}^n \left( \frac{\min(N_1, M_2)}{\min(N_2, M_2)} - 1 \right) I(\boldsymbol{s}_{i,2}; \boldsymbol{x}_{1,2}^n|\boldsymbol{s}_{i+1,2}^n, \mathbf{H}^n) + nC' \\
&\overset{(g)}{=} \left( \frac{\min(N_1, M_2)}{\min(N_2, M_2)} - 1 \right) I(\boldsymbol{s}_{1,2}^n; \boldsymbol{x}_{1,2}^n|\mathbf{H}^n) + nC',
\end{aligned}$$

where $(a)$ can be obtain by rearranging the summation; $(b)$ and $(g)$ are due to chain rule; $(c)$ and $(f)$ are due to the fact that $(\boldsymbol{s}_{i,1}, \boldsymbol{s}_{i,2})$—$\boldsymbol{x}_{i,2}$—$(\boldsymbol{x}_{1,2}^{i-1}, \boldsymbol{x}_{i+1,2}^n)$ is Markov; $(d)$ applies Lemma 3; and $(e)$ follows from the fact that $\boldsymbol{s}_{i,2}$—$\boldsymbol{x}_{i,2}$—$\boldsymbol{s}_{1,1}^{i-1}$ is Markov. Therefore, we have established

$$I(\boldsymbol{s}_{1,1}^n; \boldsymbol{x}_{1,2}^n|\mathbf{H}^n) \ge \frac{\min(N_1, M_2)}{\min(N_2, M_2)} I(\boldsymbol{s}_{1,2}^n; \boldsymbol{x}_{1,2}^n|\mathbf{H}^n) + nC' \quad (23)$$

Now, multiplying two sides of (11) with $\frac{\min(N_1, M_2)}{\min(N_2, M_2)}$ and adding it with (11), we have

$$\begin{aligned}
&R_1 + \frac{\min(N_1, M_2)}{\min(N_2, M_2)} R_2 \le \frac{1}{n} I(\boldsymbol{y}_{1,1}^n; \boldsymbol{x}_{1,1}^n, \boldsymbol{x}_{1,2}^n|\mathbf{H}^n) \\
&- \frac{1}{n} I(\boldsymbol{s}_{1,1}^n; \boldsymbol{x}_{1,2}^n|\mathbf{H}^n) + \frac{1}{n} \frac{\min(N_1, M_2)}{\min(N_2, M_2)} I(\boldsymbol{s}_{1,2}^n; \boldsymbol{x}_{1,2}^n|\mathbf{H}^n) \\
&\quad + \left( \frac{\min(N_1, M_2)}{\min(N_2, M_2)} + 1 \right) \epsilon(n). \qquad (24)
\end{aligned}$$

Substituting (23) into (24), we have

$$\begin{aligned}
&R_1 + \frac{\min(N_1, M_2)}{\min(N_2, M_2)} R_2 \le \frac{1}{n} I(\boldsymbol{y}_{1,1}^n; \boldsymbol{x}_{1,1}^n, \boldsymbol{x}_{1,2}^n|\mathbf{H}^n) - C' \\
&\quad + \left( \frac{\min(N_1, M_2)}{\min(N_2, M_2)} + 1 \right) \epsilon(n). \qquad (25)
\end{aligned}$$

Note that $I(\boldsymbol{y}_{1,1}^n; \boldsymbol{x}_{1,1}^n, \boldsymbol{x}_{1,2}^n|\mathbf{H}^n)$ can be viewed in terms of sum rate of the MAC channel [21]. Therefore, letting $n \to \infty$, we have

$$\begin{aligned}
&R_1 + \frac{\min(N_1, M_2)}{\min(N_2, M_2)} R_2 \\
&\le \mathbb{E}\left[\log\det\left(\mathrm{I} + \frac{P}{\sigma^2 M_1}\mathbf{H}_{11}\mathbf{H}_{11}^\dagger + \frac{P}{\sigma^2 M_2}\mathbf{H}_{21}\mathbf{H}_{21}^\dagger\right)\right] + C'. \qquad (26)
\end{aligned}$$

Note that [22]

$$\lim_{P\to\infty} \frac{\log\left(\det\left(\mathrm{I} + \frac{P}{\sigma^2 M_1}\mathbf{H}_{11}[1]\mathbf{H}_{11}^\dagger[1] + \frac{P}{\sigma^2 M_2}\mathbf{H}_{21}[1]\mathbf{H}_{21}^\dagger[1]\right)\right)}{\log\left(1 + \frac{P}{\sigma^2}\right)} = \min(M_1 + M_2, N_1).$$

Therefore, we complete the proof by dividing both sides of (26) with $\log\left(1 + P/\sigma^2\right)$ and letting $P \to \infty$.

## V. Conclusion

In this paper, we derive an outer bound for the DoF region of two-user MIMO interference isotropic fading channel with channel state known at receivers only. The results demonstrate the loss of DoF due to the lack of CSI at transmitters, which suggests that CSI at transmitters is crucial for improving the capacity at high SNRs. Techniques for acquiring CSI at transmitters, such as CSI feedback and user coordination requires further investigation. An interesting extension of this work is to study the region of the generalized DoFs defined in [1]. Finally, it has come to our attention that in a very recent work [23], Huang *et al.* have obtained essentially the same DoF region in the case of Rayleigh fading, which is a special form of isotropic fading considered in this paper.

## Appendix I
## Proof of the Redundancy of (5) with Respect to (4)

Denote the intersections of line

$$d_1 + \frac{\min(N_1, M_2)}{\min(N_2, M_2)} d_2 = \min(M_1 + M_2, N_1) \quad (27)$$

with $d_1$-axis and $d_2$-axis by $(a_1, 0)$ and $(0, b_1)$, respectively, and denote the intersections of line

$$d_1 + d_2 = \min(M_1 + M_2, N_2)\,. \quad (28)$$

with $d_1$-axis and $d_2$-axis by $(a_2, 0)$ and $(0, b_2)$, respectively. Geometrically, it suffices to show that $a_2 \geq a_1$ and $b_2 \geq b_1$.

Note that $a_i = \min(M_1 + M_2, N_i)$. With $N_2 \geq N_1$, we have $a_2 \geq a_1$. The remaining is to evaluate $b_1$ and $b_2$ case by case. If $N_2 \geq N_1 \geq M_2$, $b_1 = \min(M_1 + M_2, N_1)$ and $b_2 = \min(M_1 + M_2, N_2)$; if $N_2 \geq M_2 > N_1$, $b_1 = M_2$ and $b_2 = \min(M_1 + M_2, N_2)$; if $M_2 > N_2 \geq N_1$, $b_1 = b_2 = N_2$. Therefore, we always have $b_2 \geq b_1$ since $N_2 \geq N_1$.

## Appendix II
## Proof of Theorem 2

Define

$$\begin{aligned}&\mathsf{mmse}(\boldsymbol{u}, \gamma)\\ &= \mathbb{E}\left[[\boldsymbol{u} - \mathbb{E}[\boldsymbol{u}|\sqrt{\gamma}\,\boldsymbol{u} + \boldsymbol{z}]]^\dagger [\boldsymbol{u} - \mathbb{E}[\boldsymbol{u}|\sqrt{\gamma}\,\boldsymbol{u} + \boldsymbol{z}]]\right] \quad (29)\end{aligned}$$

where $\boldsymbol{z} \sim \mathcal{CN}(0, \sigma^2 \mathrm{I})$ and is independent of $\boldsymbol{u}$.

Since $\mathrm{B}\mathrm{B}^\dagger = \mathrm{I}$, we have

$$\mathsf{mmse}(\mathrm{B}_i \boldsymbol{x}, \gamma) = \mathbb{E}\left[\left\|\mathrm{B}_i \boldsymbol{x} - \mathbb{E}[\mathrm{B}_i \boldsymbol{x}|\sqrt{\gamma}\,\mathrm{B}_i \boldsymbol{x} + \mathrm{B}_i \boldsymbol{z}]\right\|^2\right].$$

Since $\mathrm{B}_i(\sqrt{\gamma}\,\boldsymbol{x} + \boldsymbol{z})$—$\sqrt{\gamma}\,\boldsymbol{x} + \boldsymbol{z}$—$\mathrm{B}_i \boldsymbol{x}$ forms a Markov Chain,

$$\begin{aligned}&\mathsf{mmse}(\mathrm{B}_i \boldsymbol{x}, \gamma)\\ &\geq \mathbb{E}\left[[\mathrm{B}_i \boldsymbol{x} - \mathbb{E}[\mathrm{B}_i \boldsymbol{x}|\sqrt{\gamma}\,\boldsymbol{x} + \boldsymbol{z}]]^\dagger [\mathrm{B}_i \boldsymbol{x} - \mathbb{E}[\mathrm{B}_i \boldsymbol{x}|\sqrt{\gamma}\,\boldsymbol{x} + \boldsymbol{z}]]\right]\\ &= \mathbb{E}\left[[\boldsymbol{x} - \mathbb{E}[\boldsymbol{x}|\sqrt{\gamma}\,\boldsymbol{x} + \boldsymbol{z}]]^\dagger \mathrm{B}_i^\dagger \mathrm{B}_i [\boldsymbol{x} - \mathbb{E}[\boldsymbol{x}|\sqrt{\gamma}\,\boldsymbol{x} + \boldsymbol{z}]]\right]\end{aligned}$$

By the geometric assumption,

$$\begin{aligned}&\sum_{i=1}^{m} p_i \mathsf{mmse}(\mathrm{B}_i \boldsymbol{x}, \gamma)\\ &\geq \sum_{i=1} p_i \mathbb{E}\left[[\boldsymbol{x} - \mathbb{E}[\boldsymbol{x}|\sqrt{\gamma}\,\boldsymbol{x} + \boldsymbol{z}]]^\dagger \mathrm{B}_i^\dagger \mathrm{B}_i [\boldsymbol{x} - \mathbb{E}[\boldsymbol{x}|\sqrt{\gamma}\,\boldsymbol{x} + \boldsymbol{z}]]\right]\\ &= \mathbb{E}\left[[\boldsymbol{x} - \mathbb{E}[\boldsymbol{x}|\sqrt{\gamma}\,\boldsymbol{x} + \boldsymbol{z}]]^\dagger [\boldsymbol{x} - \mathbb{E}[\boldsymbol{x}|\sqrt{\gamma}\,\boldsymbol{x} + \boldsymbol{z}]]\right]\\ &= \mathsf{mmse}(\boldsymbol{x}, \gamma) \quad (30)\end{aligned}$$

Integrating over $\gamma$ from 0 to 1, (16) follows from the relation between the MMSE and the mutual information [20].